\documentclass[a4paper]{jpconf}
\usepackage{graphicx}
\begin{document}
\title{High Energy Cosmic Ray Interactions -- an Overview}

\author{Sergey Ostapchenko}

\address{Forschungszentrum Karlsruhe, Institut f\"ur Kernphysik, 76021 Karlsruhe, Germany}
\address{D.V. Skobeltsyn Institute of Nuclear Physics, Moscow
State University, 119992 Moscow, Russia}

\ead{serguei@ik.fzk.de}

\begin{abstract}
The status of present theoretical description of very high energy hadronic
interactions is reviewed. The impact of new results of accelerator 
and cosmic ray experiments on hadronic interaction model constructions is 
discussed in detail. Special attention is payed to remaining uncertainties
in model extrapolations into the ultra-high energy domain, in particular,
concerning model predictions for the muon component of extensive air showers.	
New promising theoretical approaches are outlined and future
experimental prospects are discussed.      
\end{abstract}

\section{Introduction}
The basic method of studying very high energy cosmic rays (CR) is the air shower technique,
which amounts to investigate nuclear-electro-magnetic cascades -- extensive air showers (EAS) --
induced by energetic cosmic ray particles in the atmosphere. Contemporary EAS experiments resemble
to a large extent their accelerator counterparts in the sense that in both cases one applies
sophisticated simulation procedures to obtain a good understanding of the measurements and to
interprete the data. In CR experiments an unavoidable complication is that one deals with 
particle interactions at energies significantly in excess of those of present colliders.
In particular, this is a severe problem for the description of the backbone of air showers -- 
the hadronic cascade, as  the corresponding microscopic theory -- 
perturbative Quantum Chromodynamics (pQCD) -- is inapplicable for the treatment of general hadronic collisions
and the data on secondary particle production in the fragmentation region, being of crucial importance
 for the description of EAS development, are rather scarce, especially, for scattering on nuclear targets. 
The situation can be improved by employing phenomenological hadronic Monte Carlo (MC) models,
 e.g.~DPMJET \cite{ran95,boo04}, QGSJET  \cite{kal94,kal97}, SIBYLL  \cite{fle94,eng99},
or the new  QGSJET-II  \cite{ost05,ost06} and  EPOS \cite{wer06,wer07} models.
 Being based on some underlying theoretical approaches, they relate different interaction characteristics to each other
and allow one to extrapolate the interaction picture to different kinematic regions and for different
types of colliding particles, within particular model framework. Correspondingly, they are characterized
by a restricted number of adjustable parameters, which can be fitted with available data.
However, any model is only a model, its microscopic content being restricted by only a number of possible
physics mechanisms. Thus, one can not exclude the possibility that something important is missing, especially,
concerning the very high energy range. On the other hand, even the underlying theoretical ideas may appear
to be wrong. This explains the need for alternative model approaches and for continuing tests of model
validity, using both accelerator and cosmic ray data. Meanwhile, the spread in  model predictions
 may give some  feeling on the corresponding uncertainties \cite{hec97}.

\section{Basic physics}
It is presently commonly accepted that high energy hadronic (nuclear) collisions are mediated by multiple
parton cascades, proceeding in parallel. At not too high energies all parton branchings in a cascade are
characterized by a small 4-momentum squared $q^2$ transferred, the processes generally referred to as 
non-perturbative 'soft' ones. By the uncertainty principle, at each step the newly emitted parton is 
separated from its parent by a non-small distance $\Delta b^2\sim 1/q^2$ in the transverse plane. 
As a consequence, the  interaction region widens with energy,
 remaining at the same moment rather dilute, with a small parton density 
per unit area. On the other hand, with the energy increasing one observes a sizable contribution 
of so-called 'semi-hard' processes, where some parton emissions proceed with a comparatively large momentum
transfer $q^2>Q_0^2$, $Q_0^2$ being some cutoff for pQCD being applicable, 
and result in the production of observed  hadron jets of high $p_t$. The  smallness of
the strong coupling $\alpha_s(q^2)$, characteristic for such processes, is compensated by large logarithmic 
factors and by high  number of partons in the cascade \cite{glr}. Such 'hard' parton branchings proceed with a negligible
displacement in the transverse plane,   leading to a rapid rise of parton densities. Thus, one can to a some
extent separate hadronic collisions at large and small impact parameters. The former are characterized by 
low parton densities and mainly proceed via non-perturbative soft processes. The latter are more and more
dominated with increasing energy by the contribution of semi-hard processes; one deals there with
large numbers of partons being closely packed, which results in significant non-linear effects:
different  cascades fuse together,  preventing  parton densities from further increase \cite{glr}.

How this picture transforms in the very high energy limit? There, dominant semi-hard contributions look as follows.
First, the underlying parton cascade develops in the soft low virtuality region, with  small $x$ partons diffusing
towards larger impact parameters. Then, hard parton branchings become efficient; new partons are produced without
 sizable transverse displacements, contributing to the overall parton density increase at a given point.
As the result, the 'black' high density region extends to larger and larger impact parameters. On the other hand,
the dilute  peripheral region always persists outside the 'black' one, 
being formed by purely soft parton cascading. In general, one expects that with increasing energy
the parton  density  in the 'black' region is saturated up to comparatively high values of parton
virtualities and the relevant processes can, in principle, be described within the perturbative formalism 
\cite{str06}, while the peripheral region remains governed by non-perturbative physics.
According to present  data,  
the relative sizes of the dense and the dilute areas remain comparable in the  collider energy range \cite{str06},
 the  peripheral collisions thus giving an important contribution to  observed quantities.

In reality, the discussed separation of central and peripheral collisions is rather crude, as the average parton 
densities rise gradually with decreasing impact parameter. Thus, there exists an important 'transition' region of
moderately large impact parameters, characterized by large but not yet saturated parton densities,
 where the contributions of both soft and semi-hard processes are of equal importance, and where non-linear parton
effects  provide sizable corrections. In fact, it is  the treatment of such non-linear interaction contributions
in the 'dense' and in the 'transition' regimes which is the main challenge for contemporary hadronic interaction models.

\section{Model approaches}
The QGSJET model, being based on the Pomeron phenomenology \cite{kai84}, 
describes hadronic multiple scattering processes as multiple exchanges of composite objects -- Pomerons, 
 corresponding to independent microscopic parton cascades.
 For the soft low virtuality cascades a phenomenological 'soft' Pomeron amplitude is employed, whereas
 semi-hard scattering processes are treated as exchanges of 'semi-hard Pomerons',
the latter composed of a piece of QCD parton ladder 'sandwiched' between two soft Pomerons \cite{kal94,ost02};
 the soft Pomerons and the ladder describing correspondingly the low and high virtuality parts of the underlying
parton cascade. The principal feature of  QGSJET-II   is the 
treatment of non-linear parton effects described as Pomeron-Pomeron interactions, based on all order re-summation of the  
corresponding Reggeon Field Theory  diagrams \cite{ost06,ost06a}. The basic assumption of the scheme is that
Pomeron-Pomeron coupling is dominated by non-perturbative parton processes and can be described by means of 
phenomenological  multi-Pomeron vertices. The approach allows one to obtain a consistent
description of various hadronic cross sections and structure functions, including diffractive ones,
 for a fixed, energy-independent $Q_0$-cutoff \cite{ost06}. 
Presently it is the most advanced model for the description of hadronic interactions in the peripheral 
 and the 'transition' regimes. However, in central  collisions one may expect
 sizable corrections to come from 'hard'
(high $q^2$) Pomeron-Pomeron coupling, which is neglected in QGSJET-II. A reasonable agreement
of the model predictions for  central nucleus-nucleus collisions with the data of RHIC collider 
 indicates the smallness of such effects \cite{ost05a}. Nevertheless, the
situation may change at much higher energies. 

SIBYLL~2.1  \cite{eng99} also employs Pomeron formalism for the description
of soft  processes, while semi-hard ones are treated in the framework of the mini-jet approach \cite{gai85},
 which is qualitatively similar to the above-discussed 'semi-hard Pomeron' 
scheme (the differences between the two
approaches are discussed in \cite{ost03,ost06b}). The  treatment of non-linear  effects in that model is 
essentially orthogonal to the one in QGSJET-II, being based on the parton saturation approach \cite{glr}.
 Namely, it is assumed that semi-hard processes result in the production of partons of transverse momenta
  larger than some effective energy-dependent
saturation scale, $Q^2_0=Q^2_{\rm sat}(s)$, for which the double leading-log ansatz \cite{glr} is used.
However, the correlation between actual parton densities and the saturation scale holds in the model
only in average sense,
 i.e.~it reflects the increase of the average density with energy; the same  scale is used both
for dense central and for dilute peripheral collisions. 
 On the other hand, non-linear effects are neglected for the 'soft' 
interaction component. The latter is partly cured in DPMJET-III \cite{boo04}, 
taking into consideration lowest order diagrams for Pomeron-Pomeron interactions. 

The EPOS model, being the successor to NEXUS \cite{dre01}, employs the above-discussed
soft and  semi-hard Pomeron scheme and, in contrast to all other MC generators, 
 takes into account energy-momentum correlations between multiple re-scatterings \cite{hla01} 
(see also \cite{ost03} for a qualitative discussion).
 The description of non-linear effects is based on an effective treatment of lowest order 
Pomeron-Pomeron interaction graphs, with the corresponding parameters being adjusted from comparison with RHIC data.  A big advantage of the model is an excellent calibration to available accelerator data. 
However, its extrapolation towards very high energies may  depend  on the adopted empirical 
 parameterizations for  non-linear interaction contributions.

\section{Model predictions and experimental data}
Although the spread in  contemporary  model predictions for EAS characteristics
 is much more moderate than  ten years ago \cite{hec97}, it is still quite significant. 
Among the most model-dependent quantities is the  shower maximum position, which  depends  on the 
corresponding results for  total inelastic and diffractive cross sections and for the inelasticity of
 proton-air interactions.
The predicted inelastic cross sections of different models stay in reasonable agreement at collider energy range
and sizably diverge at highest CR energies, the largest values coming from the SIBYLL model. 
On the other hand, model predictions for the inelasticity of proton-air interactions and
 for the rate of diffraction processes differ significantly at all energies. 
 In general, both cross sections and inelasticities are likely to be dominated by the contribution of 
hadronic collisions in the peripheral and the 'transition' regimes,  for which 
 QGSJET-II provides a more elaborate description, compared to other MC generators.
 An important test of model predictions will be provided by the LHCf experiment \cite{kas07},
which will measure  leading neutron spectra  in proton-proton collisions.
This observable appears to be  sensitive both to the average inelasticity and to  projectile
proton diffraction dissociation; there are almost an order of magnitude differences in model predictions 
for the forward neutron spectra, which can be well discriminated by the experiment \cite{kas07}.
 On the other hand, the high  energy  behavior of the total proton-proton cross section 
  will be reliably fixed by the corresponding LHC measurement.

An important topic are model predictions for the CR muon component. Here one has to distinguish
between the results for inclusive muon spectra and for  EAS muon content for a given primary energy.
 The former are dominated by single interactions of primary protons of  energies in average only an order
  of magnitude higher than the ones of measured muons. Due to the  steepness of the primary CR spectrum, 
  the corresponding results
are very sensitive to the shape of the forward pion and kaon spectra in proton-air collisions.
Comparing with recent  accelerator measurements \cite{na49}, one finds that the old QGSJET  predicts
 too soft  pion spectra, while  SIBYLL and QGSJET-II stay in a reasonably good agreement with data.
  However, the energy dependence is quite different in the two models:  SIBYLL predicts  rather precise 
  Feynman scaling in the energy range of interest ($0.1\div 10$ TeV), which is supported by inclusive
  muon flux measurements \cite{ung05}, whereas QGSJET-II shows a noticeable scaling violation. 

The EAS muon content is formed during a multi-step hadronic cascade process and mainly depends on the
 total  multiplicity of hadron-air collisions. The differences between SIBYLL and QGSJET-II
for the predicted $N_{\mu}$ are only at $10\div 15$\% level and from general considerations one might expect
that the old QGSJET gives the upper limit for the  EAS muon number \cite{ost06b}.
However, the EPOS predictions for $N_{\mu}$ appeared to be significantly higher than in any other model  \cite{wer07}, being in  particular almost
a factor of two in excess of those of QGSJET-II. This seems only possible if
the multiplicity of pion-air collisions in increased by a similar large factor \cite{ost06b}.
 While it is problematic to verify that in collider experiments, EPOS predictions can be confronted with
  KASCADE air shower data \cite{ant05}; if supported by KASCADE,  EPOS may lead to a serious
  revision of  present views
on  CR composition in the above-knee energy range.

\smallskip
{\small Informative discussions with  R. Engel, T. Pierog, and M. Strikman are highly acknowledged.}

\end{document}